\documentclass[aps,amsmath, prb,twocolumn,superscriptaddress]{revtex4-2}
\usepackage{lineno}
\usepackage{graphicx}
\usepackage{dcolumn}
\usepackage{bm}
\usepackage{lipsum}
\usepackage{color}
\usepackage{amsfonts,eucal,mathrsfs,amssymb}
\usepackage[table]{xcolor}
\usepackage{multirow}

\renewcommand{\vec}[1]{\bm{#1}}
\newcommand{\dd}{\mathrm{d}}

\usepackage[breaklinks=true,unicode=true,urlcolor = blue,colorlinks = true,citecolor = blue,linkcolor = blue]{hyperref}

\begin{document}

\preprint{}
\title{Curvature-induced skyrmion deformation}

\author{Artur Bichs}
\affiliation{Institute for Complex Quantum Systems, Ulm University, Alber-Einstein-Allee 11, D-89069, Germany}
\affiliation{Institut f{\"u}r Theoretische Festk{\"o}rperphysik, Karlsruher Institut für Technologie, 76131 Karlsruhe, Germany}

\author{Kostiantyn V. Yershov}
\affiliation{Institute for Theoretical Solid State Physics, Leibniz Institute for Solid State and Materials Research Dresden, Helmholtzstr. 20, D-01069 Dresden, Germany}
\affiliation{Bogolyubov Institute for Theoretical Physics of the National Academy of Sciences of Ukraine, 03143 Kyiv, Ukraine}

\author{Volodymyr P. Kravchuk}
\affiliation{Institute for Theoretical Solid State Physics, Leibniz Institute for Solid State and Materials Research Dresden, Helmholtzstr. 20, D-01069 Dresden, Germany}
\affiliation{Bogolyubov Institute for Theoretical Physics of the National Academy of Sciences of Ukraine, 03143 Kyiv, Ukraine}

%
%


\begin{abstract}
Here we perform a systematic analysis of equilibrium solutions for N{\'e}el and Bloch skyrmions on a cylindrical shell and demonstrate that both experience curvature-induced deformation. Considering the curvature as a small perturbation and applying the first-order perturbation theory we decompose the deformation on the radial symmetrical and elliptical parts. The orientation of the elliptically deformed skyrmion shape depends on the skyrmion type. For N{\'e}el skyrmion, the long semi-axis of the ellipse is oriented along or perpendicularly to the cylinder generatrix depending on the signs of the curvature and the Dzyaloshinskii--Moriya interaction. In contrast, the elliptical shape of the Bloch skyrmion makes an angle $\pm\pi/4$ with the generatrix. The theoretically predicted skyrmion shapes are compared to those obtained from the numerical spin-lattice simulations. Using simulations we also investigate the nonlinear regime beyond applicability of the first-order perturbation theory. In particular, we demonstrate that N{\'e}el skyrmion can collapse for a large enough curvature.
\end{abstract}

\maketitle
\section{Introduction}
Magnetic skyrmions are topological solitons stabilized in noncentrosymmetric magnets due to the Dzyaloshinskii--Moriya interaction (DMI)~\cite{Seki16,Liu20c,Back20,Nagaosa13,Fert17,Wiesendanger16,Bogdanov20a}.  The topological protection and the resistance to external influences (e.g. temperature, external magnetic field), make skyrmions promising elements for transmitting and storing information in spintronic devices~\cite{Sampaio13,Tomasello14,Zhang15b,Zhang15c}. The transition from two-dimensional to three-dimensional architecture of spintronic devices allows for a significant increase in information capacity and processing speed~\cite{Parkin08,Fernandez17,Fischer20}. Since the three-dimensional devices unavoidably contain curvilinear elements, the influence of curvature on the properties of magnetic skyrmions must be considered. The number of new curvature-induced features of skyrmions has already been found. It was shown that skyrmion can be stabilized on the curvilinear shell without intrinsic DMI~\cite{Kravchuk16a}. Local curvilinear defect (bump) of the magnetic film may create pinning, as well as depinning potential for skyrmion~\cite{Kravchuk18a,Pylypovskyi18a,Carvalho-Santos20,Carvalho-Santos21}. The bumps arranged in the periodical array may induce the skyrmion lattice as a ground state~\cite{Kravchuk18a}. Such a kind of skyrmion lattice does not require an external magnetic field and can be of arbitrary symmetry. Skyrmion pinned on a large-amplitude bump can demonstrate a multiplet of states differing in the radius of the skyrmion~\cite{Kravchuk18a}. Curvature influences skyrmion dynamics. The spectrum of the linear magnon excitations of the skyrmion pinned on the bump is significantly modified because of the curvature~\cite{Korniienko20}. Due to translation symmetry breaking the translational zero mode is transformed into the low-frequency gyromode which corresponds to the uniform rotation of the skyrmion center around the bump extremum point~\cite{Korniienko20}. The gradient of the mean curvature plays the role of the driving force acting on the skyrmion resulting in the skyrmion drift along the curvilinear surface~\cite{Yershov22}. Curvature can significantly increase the skyrmion stability against the temperature fluctuations~\cite{Silva-Junior24}.

Recently we demonstrated that the curvature-induced deformation of skyrmions is crucial for describing the skyrmion drift induced by the curvature gradients~\cite{Yershov22}. In contrast to Ref.~\onlinecite{Yershov22}, where the skyrmion deformation was described phenomenologically by means of the collective variables method, here we obtain the exact solution for the deformed skyrmions within the first-order perturbation theory considering curvature as a small perturbation.

Here we limit ourselves to the simplest cylindrical geometry which, however, is experimentally accesible~\cite{Krusin-Elbaum04,Lenz19,Kaniukov18,Pereira18,Koerber21a} and manifests a rich spectra of the curvature-related effects in skyrmion dynamics, namely the high-speed propagation of current-driven skyrmions~\cite{Wang19}, skyrmion Walker breakdown in perpendicular field~\cite{Wang23}, skyrmion dynamics driven by rotary magnetic field~\cite{Chi21}, formation and annihilation of skyrmions in a bucket-shaped nanotube~\cite{Yu21a}.

\section{Model}\label{sec:model}

\begin{figure*}
    \centering
    \includegraphics[width=\linewidth]{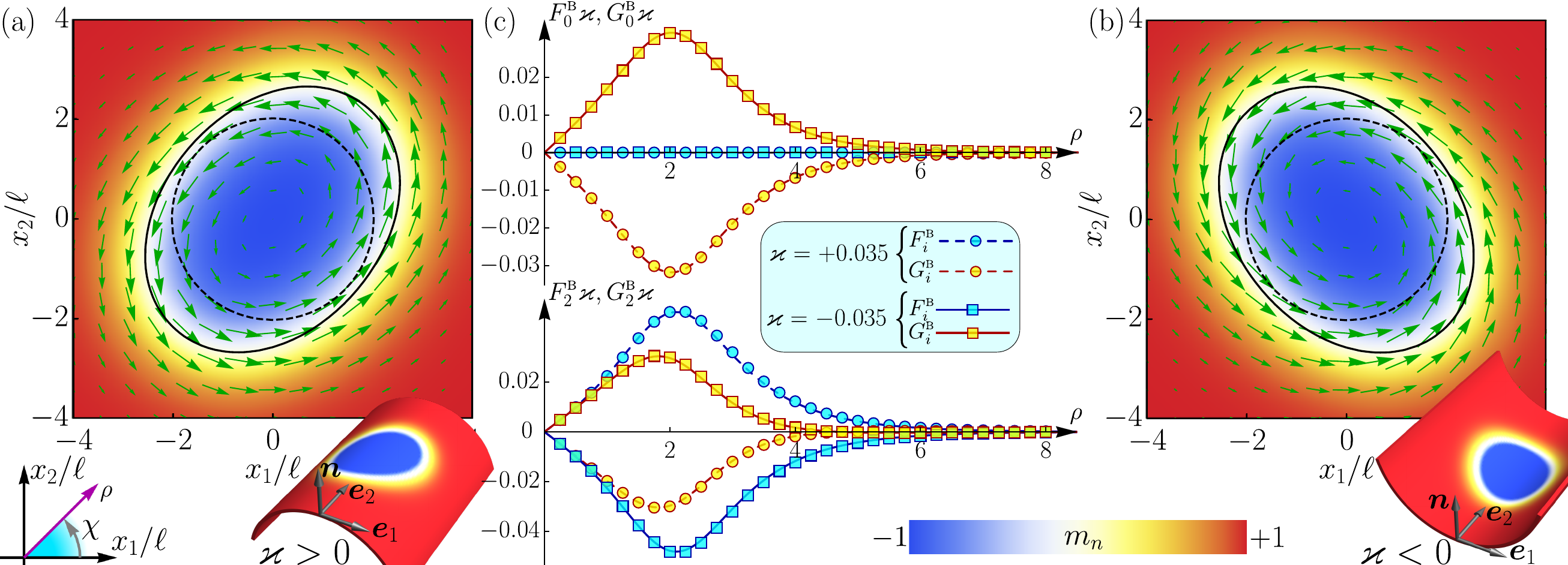}
    \caption{Deformation of a Bloch skyrmion on a cylindrical surface is shown for a positive $\varkappa=+0.175$ (a) and negative $\varkappa=-0.175$ (b) curvatures. In both cases $d=1.1$. The tangent $\vec{\mu}$ and normal $m_n$ magnetization components are shown by arrows and color gradients, respectively. The skyrmion shape is also shown by the solid and dashed isolines $m_n=0$ which correspond to the deformed ($\varkappa\ne0$) and planar ($\varkappa=0$) cases. The data for pannels (a) and (b) are obtained by means of the spin-lattice simulations. Panel (c) compares the analytically obtained deformation amplitudes~\eqref{eq:expansion} (lines) with thoes extracted from the simulations (markers), for details see Appendix~\ref{app:sim}.}
    \label{fig:deform-bloch}
\end{figure*}

Let us start with a minimal model for easy-axial magnets that allows skyrmion solutions~\cite{Bogdanov94,Rohart13,Komineas15c,Leonov16,Kravchuk18}. The model Hamiltonian
\begin{equation}\label{eq:model}
    \mathcal{H}=h\int\left[A\mathscr{E}_{ex}+K\mathscr{E}_{an}+D\mathscr{E}_{\textsc{dmi}}\right]\dd x_1\dd x_2
\end{equation}
takes into account three interactions, namely the isotopic exchange with $\mathscr{E}_{ex}=-\vec{m}\cdot\nabla^2\vec{m}$ where $\vec{m}$ is the unit magnetization vector, the easy-normal anisotropy with $\mathscr{E}_{an}=1-m_n^2$ and $K>0$ where $m_n$ is the magnetization component normal to the surface, and DMI represented by $\mathscr{E}_{\textsc{dmi}}$. We consider two cases: the bulk DMI with $\mathscr{E}_{\textsc{dmi}}^\textsc{b}=\vec{m}\cdot[\vec{\nabla}\times\vec{m}]$  and the interfacial DMI $\mathscr{E}_{\textsc{dmi}}^\textsc{i}=m_n(\vec{\nabla}\cdot\vec{m})-\vec{m}\cdot\vec{\nabla}m_n$. For a planar film, Bloch and N{\'e}el skyrmions are stabilizing  for the cases $\mathscr{E}_{\textsc{dmi}}=\mathscr{E}_{\textsc{dmi}}^{\textsc{b}}$ and $\mathscr{E}_{\textsc{dmi}}=\mathscr{E}_{\textsc{dmi}}^{\textsc{i}}$, respectively.  We assume that the film thickness $h$ is small enough to ensure the magnetization uniformity along the normal direction. Model~\eqref{eq:model} determines typical length scale $\ell=\sqrt{A/K}$ which is a domain wall thickness. Using $\ell$ as a length unit, one obtains that the  dimensionless DMI constant $d=D/\sqrt{AK}$ is the only control parameter of the model~\eqref{eq:model}. For a planar film and $|d|<d_c=4/\pi$, the ground state of Hamiltonian \eqref{eq:model}  is the uniform magnetization perpendicular to the film and skyrmions exist as topological excitations of this uniform ground state~\cite{Bogdanov94,Rohart13,Komineas15c,Leonov16,Kravchuk18}. When $|d|$ approaches its critical value $d_c$, the skyrmion radius diverges~\cite{Rohart13,Kravchuk18}, and for $|d|>d_c$ the ground state changes to a peridical structure of the magnetization.

\begin{figure*}
    \centering
    \includegraphics[width=\linewidth]{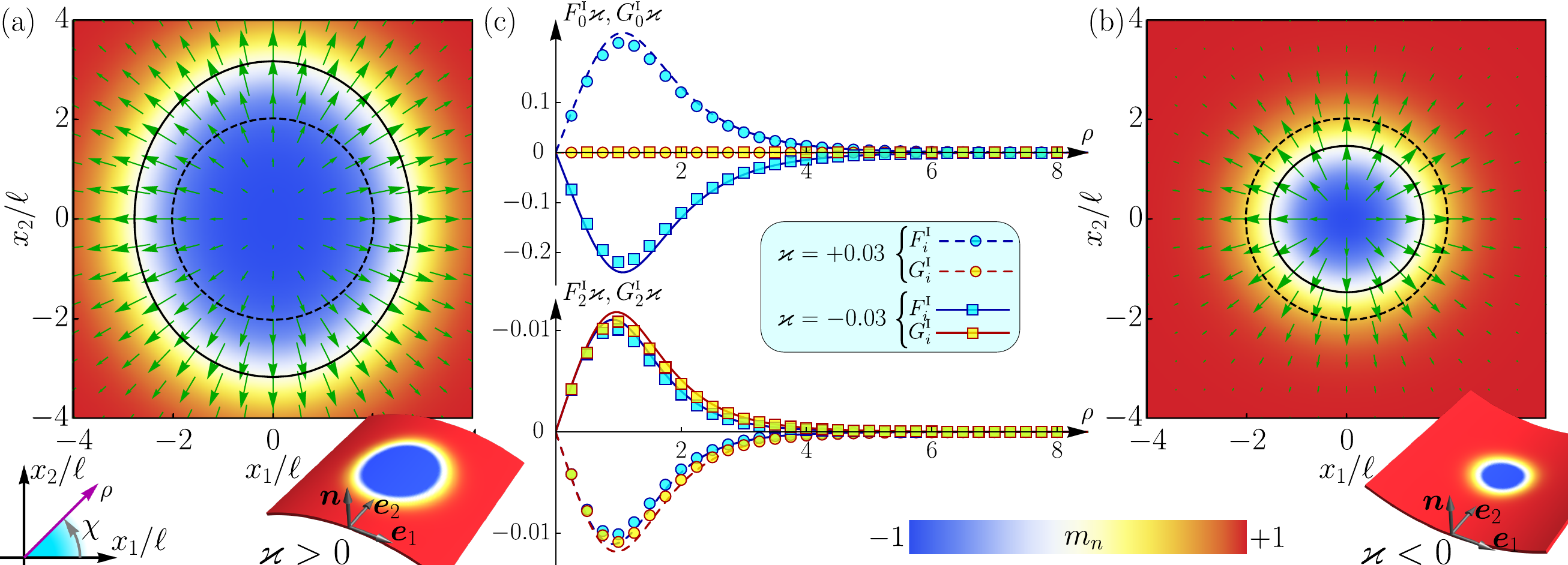}
    \caption{Deformation of a N{\'e}el skyrmion on a cylindrical surface is shown for a positive $\varkappa=+0.06$ (a) and negative $\varkappa=-0.06$ (b) curvatures. $d=1.1$ for panels (a,b) and $d=0.9$ for panel (c). The rest of notations are the same as in Fig.~\ref{fig:deform-bloch}. 
    }
    \label{fig:deform-neel}
\end{figure*}

Our interest is to study the influence of the film's cylindrical deformation on skyrmion solutions. On the surface of the cylinder, we introduce curvilinear coordinates $x_1$ and $x_2$, which are the arc lengths along the cylinder directrix and generatrix, respectively. An advantage of the introduced coordinates is the Eucledian metric tensor $g_{ij}=\text{diag}(1,1)$. 

Previously we established the dependence of the critical DMI constant $d_c$ on the diectrix curvature $\kappa=\pm 1/R$ with $R$ being the cylinder radius~\cite{Yershov20}. Here we figure out the impact of $\kappa$ on the skyrmion shape. The sign of $\kappa$ is defined in Fig.~\ref{fig:deform-bloch}. For our purposes, it is convenient to introduce parameterization of the unit magnetization vector by spherical angles $\vec{m}=\sin\theta\vec{\mu}+\cos\theta\vec{n}$, where $\vec{\mu}=\vec{e}_1\cos\phi+\vec{e}_2\sin\phi$ is projection of the magnetization on the tangent plane with $\vec{e}_1$ and $\vec{e}_2$ being the tangential unit basis vectors in directions of the coordinates $x_1$ and $x_2$, respectively. $\vec{n}=\vec{e}_1\times\vec{e}_2$ is the unit normal, see Fig.~\ref{fig:deform-bloch}.

For a curved surface, one presents the exchange energy in form of three contributions $\mathscr{E}_{ex}=\mathscr{E}_{ex}^0+\mathscr{E}_{ex}^{\textsc{dmi}}+\mathscr{E}_{ex}^{an}$. Here $\mathscr{E}_{ex}^0=(\vec{\vec{\nabla}}\theta)^2+\sin^2\theta(\vec{\nabla}\phi)^2$ is the common exchange. In our case, the curvature-induced DMI and anisotropy have the form $\mathscr{E}_{ex}^{\textsc{dmi}}=2\kappa(\cos\phi\partial_1\theta-\sin\phi\sin\theta\cos\theta\partial_1\phi)$ and $\mathscr{E}_{ex}^{an}=\kappa^2(1-\sin^2\theta\sin^2\phi)$, respectively. For more details, see Refs. \cite{Gaididei14,Sheka22b}. The DMI contributions are as follows $\mathscr{E}_{\textsc{dmi}}^\textsc{b}=\sin^2\theta[2(\vec{\nabla}\theta\times\vec{\mu})\cdot\vec{n}+\kappa\cos\phi\sin\phi]$~\cite{Yershov19a,Sheka22b} and $\mathscr{E}_{\textsc{dmi}}^\textsc{i}=2\sin^2\theta(\vec{\nabla}\theta\cdot\vec{\mu})+\kappa\cos^2\theta$~\cite{Kravchuk18,Sheka22b}. The anisotropy contribution is trivial: $\mathscr{E}_{an}=\sin^2\theta$.

Using that $\kappa$ is constant for the cylinder, we integrate the second summand in $\mathscr{E}_{ex}^{\textsc{dmi}}$ by parts and neglecting the boundary effects we replace $\mathscr{E}_{ex}^{\textsc{dmi}}$ by ${\mathscr{E}_{ex}^{\textsc{dmi}}}'=4\kappa\sin^2\theta\partial_1\theta\cos\phi$ in the total energy functional. Since ${\mathscr{E}_{ex}^{\textsc{dmi}}}'$ has the same structure as the curvature-independent part of $\mathscr{E}_{\textsc{dmi}}^\textsc{i}$ (the first summand), we conclude that the curvature-induced effective DMI is of the interfacial type for the cylindrical geometry. Thus, we expect a stronger impact of curvature on the Néel skyrmion than on the Bloch skyrmion, because of the direct competition of the intrincic and the curvature-induced DMIs. The latter effect was already indicated in Refs.~\onlinecite{Yershov22,Korniienko20}.

\section{Skyrmion deformation}
Let us first formulate the exact equations for an equilibrium magnetization state on a cylindrical surface. To this end, the folloving change of variables $x_1=\ell\rho\cos\chi$ and $x_2=\ell\rho\sin\chi$ is convenient. Here $\rho$ and $\chi$ are the dimensionless radial distance and azimuth of the polar reference frame drawn on the cylinder surface. The Euler-Lagrange equations $\delta\mathcal{H}/\delta\theta=0$ and $\delta\mathcal{H}/\delta\phi=0$ obtain the following explicit form
\begin{subequations}\label{eq:theta-phi}
\begin{align}
    \label{eq:theta}\nonumber-\nabla^2\theta&+\sin\theta\cos\theta\left[1+(\nabla\phi)^2-\varkappa^2\sin^2\phi\right]\\
    &+2\varkappa\sin^2\theta\sin\phi\,\tilde{\partial}_1\phi+d\,\Xi^{\alpha}_{\theta}=0,\\
   \label{eq:phi} \nonumber-\vec{\nabla}\cdot&\left(\sin^2\theta\vec{\nabla}\phi\right)-2\varkappa\sin^2\theta\sin\phi\tilde{\partial}_1\theta\\
   &-\varkappa^2\sin^2\theta\sin\phi\cos\phi+d\,\Xi^{\alpha}_{\phi}=0.
    \end{align}
\end{subequations}
Here $\varkappa=\ell\kappa$ is the dimensionless curvature, $\tilde{\partial}_1=\ell\partial_1=\cos\chi\partial_\rho-\rho^{-1}\sin\chi\partial_\chi$, and the form of blocks $\Xi^\alpha_{\theta,\phi}$, with $\alpha\in\{\textsc{b},\textsc{i}\}$, depends on the DMI type. Namely, $\Xi^{\textsc{b}}_{\theta}=-\sin^2\theta\,\hat{\mathfrak{D}}^{\textsc{b}}_\phi\phi+\frac{\varkappa}{4}\sin2\theta\sin2\phi$ and $\Xi^{\textsc{b}}_{\phi}=\sin^2\theta(\hat{\mathfrak{D}}^{\textsc{b}}_\phi\theta+\frac{\varkappa}{2}\cos2\phi)$ for the bulk DMI. While $\Xi^{\textsc{i}}_{\theta}=-\sin^2\theta\,\hat{\mathfrak{D}}^{\textsc{i}}_\phi\phi-\varkappa\sin\theta\cos\theta$ and $\Xi^{\textsc{i}}_{\phi}=\sin^2\theta\,\hat{\mathfrak{D}}^{\textsc{i}}_\phi\theta$ for the interfacial DMI. Here we introduced the following differential operators $\hat{\mathfrak{D}}^{\textsc{b}}_\phi=\cos(\phi-\chi)\partial_\rho+\rho^{-1}\sin(\phi-\chi)\partial_\chi$, and $\hat{\mathfrak{D}}^{\textsc{i}}_\phi=\hat{\mathfrak{D}}^{\textsc{b}}_{\phi+\pi/2}$.


For the planar case ($\varkappa=0$), the solution of Eq.~\eqref{eq:phi} which minimizes energy \eqref{eq:model} is $\phi=\Phi^{\textsc{b}}=\chi+\varsigma\pi/2$ and $\phi=\Phi^{\textsc{i}}=\chi+(1-\varsigma)\pi/2$ for the bulk and interfacial DMI, respectively. Here $\varsigma=\text{sign}(d)=\pm1$. In both cases, $\theta=\Theta(\rho)$ with the boundary conditions $\Theta(0)=\pi$ and $\Theta(\infty)=0$. The profile of the planar skyrmion $\Theta(\rho)$ is universal for both types of DMI and is determined by Eq.~\eqref{eq:theta}, namely~\cite{Leonov16,Bogdanov89r,Bogdanov94,Rohart13,Komineas15c,Kravchuk18}
\begin{equation}
    \label{eq:Theta}
    -\nabla_\rho^2\Theta+\sin\Theta\cos\Theta\left(1+\frac{1}{\rho^2}\right)-\frac{|d|}{\rho}\sin^2\Theta=0,
\end{equation}
where $\nabla_\rho^2f=\rho^{-1}\partial_\rho(\rho\partial_\rho f)$ is the radial part of the Laplace operator. 

In the next step, we consider curvature as a small perturbation resulting in small deviations from the planar skyrmion profile: $\theta=\Theta+\vartheta$ and $\phi=\Phi^\alpha+\varphi/\sin\Theta$, where the deviations $\vartheta$ and $\varphi$ are assumed to be of the order of $\varkappa$. The linearization of Eqs.~\eqref{eq:theta-phi} with respect to $\vartheta$, $\varphi$ and $\varkappa$ results in
\begin{equation}\label{eq:lin}
    \begin{bmatrix}
        -\nabla^2+U_1 & W\partial_\chi \\
        -W\partial_\chi & -\nabla^2+U_2
    \end{bmatrix}\vec{\Psi}=\varkappa\varsigma\left[\vec{\xi}^\alpha_0(\rho)+\mathbb{M}^\alpha(\chi)\vec{\xi}^\alpha_2(\rho)\right],
\end{equation}
where $\vec{\Psi}=(\vartheta,\varphi)^{T}$. While the left-hand side of Eq.~\eqref{eq:lin} with the potentials~\cite{Kravchuk18}
\begin{equation}
    \begin{split}
        &U_1=\cos2\Theta\left(1+\frac{1}{\rho^2}\right)-\frac{|d|}{\rho}\sin2\Theta,\\
        &U_2=\cos^2\Theta\left(1+\frac{1}{\rho^2}\right)-\Theta'^2-|d|\left(\Theta'+\frac{\sin2\Theta}{2\rho}\right),\\
        &W=\frac{2}{\rho^2}\cos\Theta-\frac{|d|}{\rho}\sin\Theta.
    \end{split}
\end{equation}
is universal for both kinds of skyrmions, the right-hand side of Eq.~\eqref{eq:lin} is DMI specific: 
\begin{equation}
    \begin{split}
    &\vec{\xi}^{\textsc{b}}_0=\begin{bmatrix}
        0\\
        \Theta'\sin\Theta
    \end{bmatrix},\quad \vec{\xi}^{\textsc{b}}_2=\begin{bmatrix}
        \frac{\sin^2\Theta}{\rho}+\frac{|d|}{2}\sin\Theta\cos\Theta\\
        \sin\Theta\left(\Theta'+\frac{|d|}{2}\right)    \end{bmatrix},\\
&\vec{\xi}^{\textsc{i}}_0=\begin{bmatrix}
        \frac{\sin^2\Theta}{\rho} + |d|\sin\Theta\cos\Theta\\
       0
    \end{bmatrix},\quad \vec{\xi}^{\textsc{i}}_2=\begin{bmatrix}
        -\frac{\sin^2\Theta}{\rho}\\
       \Theta'\sin\Theta
    \end{bmatrix},
    \end{split}
\end{equation}
and $\mathbb{M}^{\textsc{b}}=\text{diag}(\sin2\chi,\cos2\chi)$, $\mathbb{M}^{\textsc{i}}=\sigma_x\mathbb{M}^{\textsc{b}}=\text{diag}(\cos2\chi,\sin2\chi)$, where $\sigma_x$ is the Pauli matrix. Here we denote $\Theta'=\partial_\rho\Theta$.
Based on the structure of the right-hand side of Eq.~\eqref{eq:lin} and on the form of $\chi$-dependednce of matrices $\mathbb{M}^\alpha$ we conclude that : (i) in the leading order in $\varkappa$ the skyrmion deformation is  a superposition of the radial-symmetrical and elliptical deformations, (ii) elliptically deformed shapes of the N{\'e}el and Bloch skyrmions are rotated at an angle $\pi/4$ relative to each other, (iii) the flip of sign of curvature is equivalent to the flip of sign of the DMI constant. Due to the latter property, without loss of generality, we fix the sign of DMI constant ($d>0$) and consider the curvature of different signs.

Equation \eqref{eq:lin} has solution $\vec{\Psi}^\alpha=\varkappa\varsigma[\vec{\Psi}^\alpha_0(\rho)+\mathbb{M}^\alpha(\chi)\vec{\Psi}^\alpha_2(\rho)]$, where the amplitudes of the radial symmetrical $\vec{\Psi}^\alpha_0$ and elliptic $\vec{\Psi}^\alpha_2$ deformations are determined by equations
\begin{equation}
    \mathbb{H}_{-\mu}\vec{\Psi}_\mu^{\textsc{b}}=\vec{\xi}_\mu^{\textsc{b}},\qquad \mathbb{H}_{\mu}\vec{\Psi}_\mu^{\textsc{i}}=\vec{\xi}_\mu^{\textsc{i}},\qquad\mu=0,2.
\end{equation}
Here 
\begin{equation}
    \mathbb{H}_\mu=\begin{bmatrix}
        -\nabla_\rho^2+\frac{\mu^2}{\rho^2}+U_1 & \mu W \\
        \mu W &  -\nabla_\rho^2+\frac{\mu^2}{\rho^2}+U_2
    \end{bmatrix}
\end{equation}
is the Hamiltonian which determines the eigenfunctions $\vec{\psi}_{\mu,\nu}$ of the magnon eigen-modes with the azimuthal quantum number $\mu$ of a planar skyrmion~\cite{Kravchuk18}: 
\begin{equation}
    \mathbb{H}_{\mu}\vec{\psi}_{\mu,\nu}=\omega_{\mu,\nu}\sigma_x\vec{\psi}_{\mu,\nu}.
\end{equation}
For each given $\mu$, operator  $\mathbb{H}_{\mu}$ generates $\nu$-spectrum of the eigenstates with eigenfrequencies $\omega_{\mu,\nu}$ and eigenfunctions $\vec{\psi}_{\mu,\nu}=(f_{\mu,\nu},g_{\mu,\nu})^{T}$. Here functions $f_{\mu,\nu}(\rho)$ and $g_{\mu,\nu}(\rho)$ determine the semiaxes of the precession ellipse of the magnetization vector $\vec{m}$ at the distance $\rho$ from the skyrmion center~\cite{Satywali21}. For each $\mu$, the eigenfunctions $\vec{\psi}_{\mu,\nu}$ compose a complete set and fulfill the orthogonality condition $\int_0^\infty\rho\,\vec{\psi}_{\mu,\nu}^T\sigma_x\vec{\psi}_{\mu,\nu'}\dd\rho=\mathcal{C}_\mu\delta_{\nu,\nu'}$ with $\mathcal{C}_\mu$ being the normalization constant~\cite{Kravchuk18}. Now we present $\vec{\Psi}_{\mu}^{\textsc{b}}$ and  $\vec{\Psi}_{\mu}^{\textsc{i}}$ in the form of series 
\begin{equation}\label{eq:expansion}
    \vec{\Psi}_{\mu}^{\textsc{b}}=\sum_{\nu}C^{\textsc{b}}_{-\mu,\nu}\vec{\psi}_{-\mu,\nu},\qquad \vec{\Psi}_{\mu}^{\textsc{i}}=\sum_{\nu}C^{\textsc{i}}_{\mu,\nu}\vec{\psi}_{\mu,\nu}
\end{equation}
and with the use of the orthogonality condition find the expansion coefficients
\begin{equation}
    C^\alpha_{\mu,\nu}=\frac{1}{\omega_{\mu,\nu}}\frac{\int_0^\infty\rho\,\vec{\psi}_{\mu,\nu}^T\vec{\xi}_\mu^\alpha\dd\rho}{\int_0^\infty\rho\,\vec{\psi}_{\mu,\nu}^T\sigma_x\vec{\psi}_{\mu,\nu}\dd\rho}.
\end{equation}
Since the eigenfunctions $\vec{\psi}_{\mu,\nu}$ are known~\cite{Kravchuk18}, one can find the amplitudes $\vec{\Psi}_\mu^\alpha$ in the form of the expansion~\eqref{eq:expansion} and reconstruct the complete deformation amplitudes $\vec{\Psi}^\alpha$. Note that the eigenfrequencies of the breathing ($\mu=0$) and elliptical ($\mu=\pm2$) do not vanish for all $\nu$~\cite{Kravchuk18} meaning that the coefficients $C_{\mu,\nu}^\alpha$ are well-defined.
In Figs.~\ref{fig:deform-bloch} and \ref{fig:deform-neel}, we compare the analytically predicted amplitudes $\vec{\Psi}_\mu^\alpha=(F_\mu^\alpha,G_\mu^\alpha)^T$ to those extracted from the spin-lattice simulations for Bloch and N{\'e}el skyrmions, respectively. 

\section{Discussion}

\begin{figure}[t]
    \includegraphics[width=0.75\columnwidth]{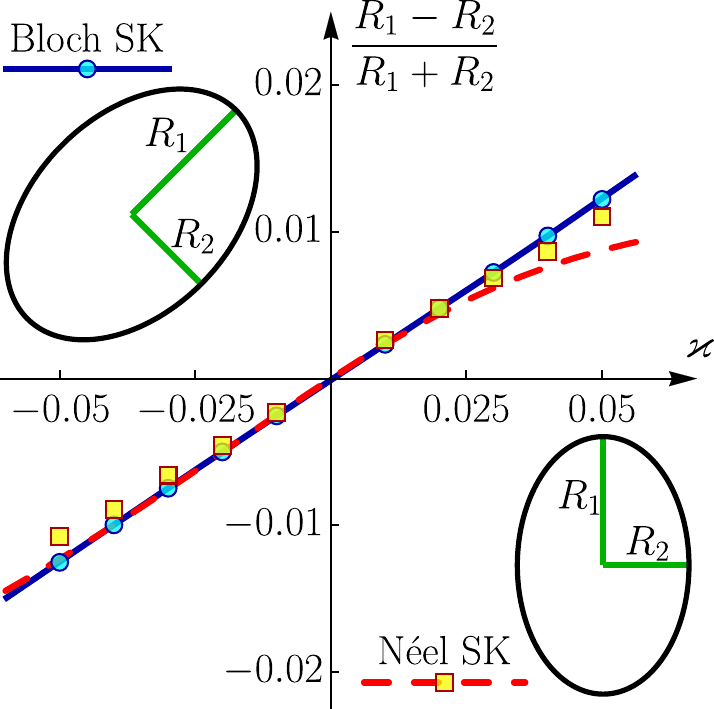}
    \caption{The skyrmion ellipticity ratio as a function of curvature for DMI $d=0.9$. Lines and markers correspond to the analytics and to the data extracted from numerical simulations, respectively. For details of numnerics see Appendix~\ref{app:sim}.}
    \label{fig:radius-ratio}
\end{figure}

Although both types of skyrmions experience radial symmetrical deformation, there are two significant differences. (i) The only tangential and only perpendicular to the surface magnetization components are affected for the Bloch and N{\'e}el skyrmions, respectively. I.e. Bloch and N{\'e}el skyrmions demonstrate the radial-symmetrical change in their helicity and radius, respectively. (ii) The deformation amplitude of the N{\'e}el skyrmion is significantly larger than those of the Bloch skyrmion. This is the consequence of the direct competition between the intrinsic and the curvature-induced DMI which is of the interfacial type for a cylinder geometry, see the discussion at the end of Section~\ref{sec:model}. Thus, the curvature effectivelly changes the constant of the interfacional DMI. On the other hand, in the subcritical regime,  when $d$ is close to the critical value $d_c$, the skyrmion radius $R_0$ is extremely sensitive to small shanges in $d$ because $R_0$ increases with $d$ faster than exponentially if $d\lessapprox d_c$~\cite{Kravchuk18}. Thus, in the subcritical regime, a small curvature results in a large change of the radius of a N{\'e}el skyrmion, making our linear theory not applicable in this case. This is the reason of using smaller $d$ for a N{\'e}el skyrmion when we compare analytical and numerical results in Fig.~\ref{fig:deform-neel}(c) and in Fig.~\ref{fig:radius-ratio}.

Both types of skyrmions experience also the elliptical deformations of comparable amplitudes (for the same $d$ and $\varkappa$). However, orientations of the elliptically deformed skyrmion cores are different, see Figs.~\ref{fig:deform-bloch},~\ref{fig:deform-neel}. They are consistent with the previous results for Bloch~\cite{Wang19,Wang23,Yershov22,Huo19} and N{\'e}el skyrmions~\cite{Yershov22}. In order to quantify the ellipticity, we introduce quantity $\mathcal{E}=(R_1-R_2)/(R_1+R_2)$, where $R_{1,2}$ are semiaxes of the elliptical skyrmion shape determined as an isoline $m_n=0$, see Fig.~\ref{fig:radius-ratio}. One can see that in the limit of small $\varkappa$, the dependence $\mathcal{E}(\varkappa)$ is linear and universal for both skyrmion types. In both cases, the change of sign of $\varkappa$ flips the sign of $\mathcal{E}$ which is equivalent to the rotation of the elliptical skyrmion shape by $\pi/2$.

\begin{figure*}
    \includegraphics[width=\linewidth]{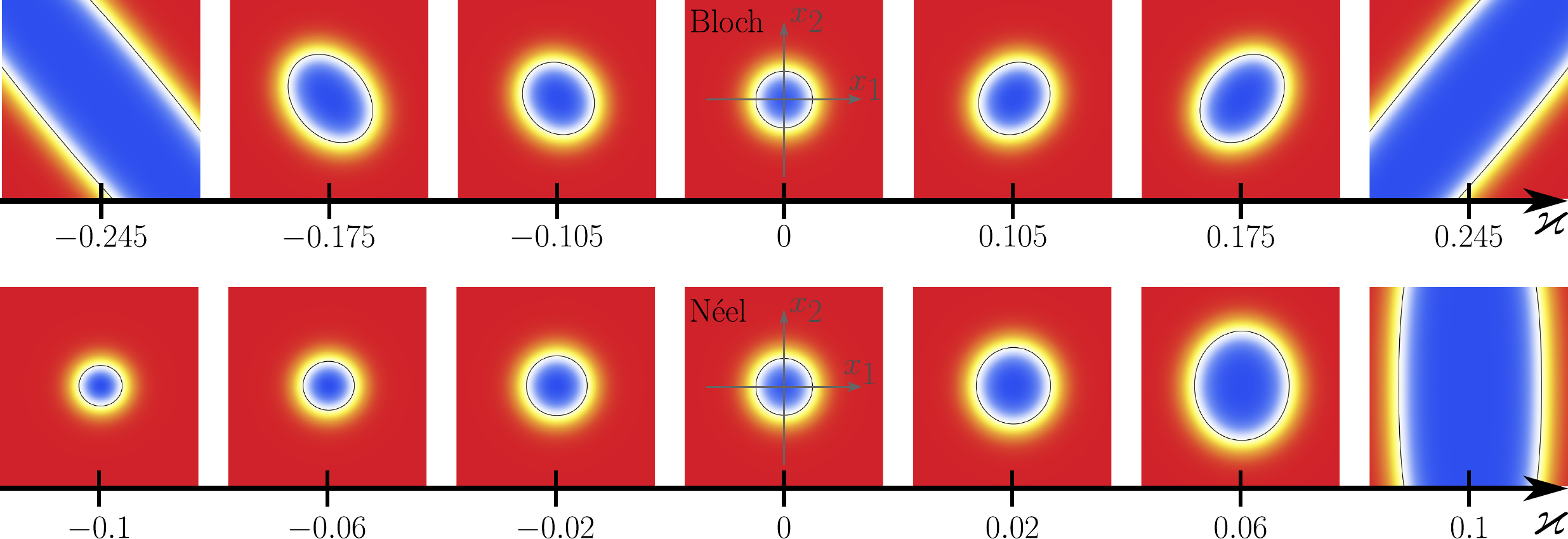}
    \caption{The evolution of Bloch (top row) and N{\'e}el (bottom row) skyrmions with curvature $\varkappa$ for $d=1.1$. The normal magnetization component $m_n$ is shown by the color sheme, which has the same meaning as in Figs.~\ref{fig:deform-bloch},~\ref{fig:deform-neel}. Each panel is of size $6\ell\times 6\ell$. The data was obtained by means of the numerical spin-lattice simulations, for details see Appendix~\ref{app:sim}.}
    \label{fig:deform-bloch-neel}
\end{figure*}

Since the presented linear theory presupposes small curvature-induced deformations, it can not be applied in the cases of large curvatures. We utilize the spin-lattice simulations in order to trace the evolution of skyrmions with curvature. The results are summarized in Fig.~\ref{fig:deform-bloch-neel}. For large curvatures, Bloch skyrmions demonstrate strong elliptical deformation and elongation in the direction which makes an angle $\text{sign}(\varkappa)\pi/4$ with the cylinder axis.  For some critical value of $|\varkappa|$, the curvature-induced transition to a stripe structure~\cite{Yershov20} takes place. The evolution of a Bloch skyrmion is the same for both signs of curvature, up to the orientation of the elongated skyrmion shape relative to the cylinder axis, see the top row in Fig.~\ref{fig:deform-bloch-neel}. This is in contrast to a N{\'e}el skyrmion whose fate depends on the curvature sign, see the bottom row in Fig.~\ref{fig:deform-bloch-neel}. For $\varkappa>0$,  the evolution of a N{\'e}el skyrmion is similar to the Bloch skyrmion. Namely, it demonstrates an elliptical elongation along the cylinder axis, and the transition to the stripe structure phase takes place for some critical value of $\varkappa$~\cite{Yershov20}. For $\varkappa<0$, a N{\'e}el skyrmion experiences the elliptical deformation in the perpendicular direction. However, it also srinks in size and collapses for some critical value of the negative curvature. This happens because of the compensation of the intrinsic interfacial DMI by the curvature-induced one. We found numerically that for $d=1.1$ the skyrmion collapces for the critical curvature within the range $-0.4<\varkappa_{\text{clps}}<-0.3$. The effect of the curvature-induced skyrmion collapse was previously predicted for spherical shells~\cite{Kravchuk16a}.

\section{Acknowledgments}
We thank Ulrike Nitzsche for technical support. This work was supported by the Deutsche Forschungsgemeinschaft (DFG, German Research Foundation) under Germany's Excellence Strategy through the W\"{u}rzburg-Dresden Cluster of Excellence on Complexity and Topology in Quantum Matter -- \emph{ct.qmat} (EXC 2147, project-ids 390858490 and 392019), Grant No. YE 232/2-1, and by the National Academy of Sciences of Ukraine (Project No. 0122U000887).

\appendix

\section{Details of numerical simulations}\label{app:sim}

Numerically, we study the skyrmion textures in cylindrical shells using an in-house developed spin-lattice simulator, for details see~\cite{Yershov22,Yershov20}.

The cylindrical surface is considered as a square lattice of size $N_1\times N_2$. We use the anisotropic Heisenberg Hamiltonian taking into account the exchange interaction, easy-normal anisotropy, and DMI. The curvature of the film enters into the problem via the coordinate-dependent normal vector $\vec{n}=\sin\left(\kappa x_1\right)\vec{e}_x + \cos\left(\kappa x_1\right)\vec{e}_z$. The system's dynamics is described by a set of $N_1 N_2$ vector Landau--Lifshitz ordinary differential equations, see  Ref.~\cite{Yershov20} for details of the cylindrical shell simulations.

To study the curvature-induced deformations of skyrmion we consider the samples of size $N_1\times N_2 = 300\times300$. The skyrmion is placed in the center of the sample. As an initial state, we use a skyrmion profile for a planar film mapped on the cylindrical shell. All simulations are performed in an overdamped regime with Gilbert damping $\alpha_\textsc{g} = 0.5$ and magnetic length $\ell = 10 a_0$ with $a_0$ being the lattice constant.

We run simulations in a long-time regime for a given curvature, type and strength of DMI to obtain an equilibrium magnetization distribution. The extraction of curvature-induced deformation amplitudes is performed in three steps. Firstly, we get the spherical angles $\theta$ and $\phi$ as functions of curvilinear coordinates: (i)~the polar angle is defined as $\theta_\text{sim} = \arccos\left(\vec{m}\cdot\vec{n}\right)$, (ii)~the azimuthal angle is defined as $\tan\phi_\text{sim} = \frac{\vec{m}\cdot\vec{e}_2}{\vec{m}\cdot\vec{e}_1}$. In the second step, we calculate curvature-induced deviations of the spherical angles as $\vartheta^\alpha\left(\rho,\chi\right) = \theta^\alpha_\text{sim}-\Theta$ and $\varphi^\alpha\left(\rho,\chi\right) = \left[\phi^\alpha_\text{sim} - \Phi^\alpha\right]\sin\Theta$ with $\alpha\in\left\{\textsc{b},\textsc{i}\right\}$. And finally, we perform a fitting procedure for curvature-induced deviations with the fitting functions $\left(\vartheta_\text{fit}^\alpha,\varphi_\text{fit}^\alpha\right)=\varkappa\left[\tilde{\vec{\Psi}}_0^\alpha + \mathbb{M}^\alpha\tilde{\vec{\Psi}}_2^\alpha\right]$, where $\tilde{\vec{\Psi}}_0$ and $\tilde{\vec{\Psi}}_2$ are the fitting parameters for each given value of $\rho$. As a result, we obtain curvature-induced deformation amplitudes $\tilde{\vec{\Psi}}_0$ and $\tilde{\vec{\Psi}}_2$ as function of $\rho$, see markers in Figs.~\ref{fig:deform-bloch}(c) and~\ref{fig:deform-neel}(c). 

The extraction of the skyrmion ellipticity ratio is performed in two steps. Firstly we get the normal component of magnetization $m_n\left(\rho,\chi\right) = \vec{m}\cdot\vec{n}$. As a second step, we calculate semiaxes $R_i$ of the elliptical skyrmion shape by solving equations: (i) $m_n^\textsc{b}(R_1,+\pi/4) = 0$ and $m_n^\textsc{b}(R_2,-\pi/4) = 0$ for Bloch skyrmions; (ii) $m_n^\textsc{i}(R_1,\pi/2) = 0$ and $m_n^\textsc{i}(R_2,0) = 0$ for N{\'e}el skyrmions. The resulting ellipticity ratio is shown in Fig.~\ref{fig:radius-ratio} by markers.


%

\end{document}